\documentclass[12pt]{article}
\input{amssym.def}
\input{amssym}
\usepackage[dvips]{color}
\usepackage{epsfig}
\usepackage{subcaption}
\usepackage{hyperref}
\normalbaselines
\makeatletter
\@addtoreset{equation}{section}
\makeatother

\def\be{\begin{equation}}
\def\ee{\end{equation}}
\def\ba{\begin{eqnarray}}
\def\ea{\end{eqnarray}}

\newcommand{\rf}[1]{(\ref{#1})}

\def\bra#1{\langle #1|}
\def\ket#1{|#1\rangle}

\def\scp{\scriptsize}

\begin{document}


\title
{Quasi-stationary states in nonlocal stochastic growth models with infinitely 
many  absorbing states}
\author{D. A. C. Jara$^1$   and   F. C. Alcaraz$^2$
\\[5mm] {\small\it
Instituto de F\'{\i}sica de S\~{a}o Carlos, Universidade de S\~{a}o Paulo, Caixa Postal 369, }\\
{\small\it 13560-590, S\~{a}o Carlos, SP, Brazil}}
\date{\today}
\maketitle
\footnotetext[1]{\tt dacj1984@gmail.com}
\footnotetext[2]{\tt alcaraz@if.sc.usp.br}

\begin{abstract}
We study a two parameter ($u,p$) extension of the conformally invariant raise
and peel model. The model  also represents a nonlocal and 
biased-asymmetric exclusion process with local and nonlocal 
jumps of excluded 
volume particles in the lattice. The model exhibits an unusual and interesting 
critical phase where, in the bulk limit, there are an infinite number of 
absorbing states. In spite of these absorbing states the system stays, during 
a time that increases exponentially with the lattice size, in a critical 
quasi-stationary state. In this critical phase the critical exponents depend 
only on  one of the parameters defining the model ($u$). The endpoint of this
 critical phase, where the system changes from an active to an inactive 
frozen phase,  belongs to a distinct universality class.
 This new 
behavior, we believe, is   due to the appearance of Jordan cells in the Hamiltonian  
describing the time evolution. The dimensions of these cells increase with the 
 lattice size. In a special case ($u=0$) where the model has no adsorptions 
we are able to calculate analytically the time evolution of some  
observables. A polynomial time dependence is obtained thanks to the 
appearance of  Jordan cells 
structures in  the Hamiltonian.

\end{abstract}

\section{ Introduction} \label{sect1}

Stochastic growth models of interfaces have been extensively studied along the 
years (see \cite{odor2008universality,hinrichsen2000non,henkel2008non} for reviews). The most studied universality class of 
critical dynamics behavior of growing interfaces are the ones represented by 
the Edward-Wilkinson (EW) \cite{EW} and the Kardar-Parisi-Zhang (KPZ) \cite{KPZ} models whose 
dynamical critical exponents are equal to 2 and $3/2$, respectively. 
Differently from these models, where the absorption and desorption processes 
are local, the raise and peel model (RPM) \cite{de2004raise}, although 
keeping the adsorption process local, the desorption processes is nonlocal. 
This model is quite interesting, as   it is the first example of an stochastic 
model with conformal invariance. The critical properties of the  model depend 
 on the parameter $u$ defined as the ratio among the adsorption and desorption 
rates. At $u=1$ the RPM is special, being exact integrable and conformally 
invariant. The dynamical critical exponent has the value $z=1$ and its 
time-evolution operator (Hamiltonian) is related to the XXZ quantum chain 
with $z$-anisotropy $\Delta = -\frac{1}{2}$  (Razumov-Stroganov 
point \cite{RS}). For $u<1$ 
(desorption rates greater than the adsorption ones) 
the model is noncritical, but 
for $u\geq1$ the model is in a critical regime with continuously varying 
critical exponents $z=z(u)$, that decreases from $z(1)=1$ (conformally 
invariant) to $z(u\to\infty)=0$.

The configurations of the growing surface in the RPM are formed by sites whose 
heights define Dyck paths  on a lattice with 
$L+1$ sites ($L$ even) and open boundaries. 
 These are staircase walks with integer horizontal steps that go from $(x,y)=(0,0)$ to 
$(L/2,L/2)$ and are never lower than the ground $x=0$ or higher than the diagonal
$x=y$\cite{wikipedia}.
 In these surface configurations there are active sites where 
adsorption and desorption processes take place, and inactive sites where nothing 
happens during the time evolution. 

An interesting extension of the RPM at 
$u=1$, proposed in \cite{alcaraz2010conformal}, is the peak adjusted raise and peel model 
(PARPM). In this model an additional parameter $p$ 
that depends on the total number of inactive sites, 
controls the relative 
changes of a given  configuration.
 The model at $p=1$ recovers the RPM. For $p\neq 1$ the model is not 
exact integrable anymore but still is conformally invariant \cite{alcaraz2010conformal}. The 
parameter $p$ in the PARPM has a limiting value ($p={p}_1=2$) where the 
configuration with only inactive sites (no adsorption or desorption) become 
an absorbing state. Surprisingly at this point, on spite of the presence of 
the absorbing state, that should be the true stationary state, the system 
stays in a quasi-stationary state during a time interval that grows 
 exponentially  
with the system size \cite{alcaraz2011conformal}. This quasi-stationary state has similar 
properties as the stationary states of the conformally invariant region 
$p<{p}_1$. 

Motivated by this unusual and interesting behavior we introduce in this paper 
an extension of the PARPM, where the parameter $p$ is extended so that when 
$p>{p}_1$ the number of absorbing states increases with the value 
of $p$. The results presented 
in this paper shows that a quasi-stationary state, with similar properties 
as in the conformally invariant region $p<{p}_1$, endures as the true 
stationary state even when the number of absorbing states is extensively 
large. 
  Only at $p=p_c>{p}_1$ the  model undergoes a transition to 
one of the infinitely many absorbing states.

\begin{figure}
\centering
\includegraphics[angle=0,width=0.4\textwidth] {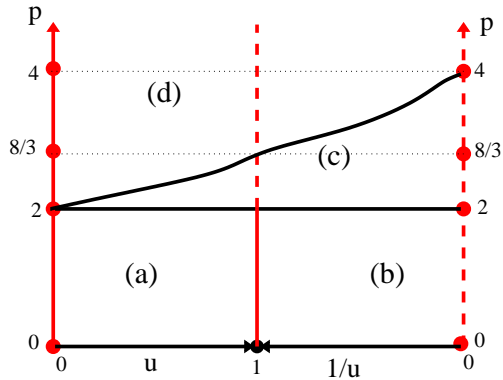}
\caption{
Schematic phase diagram of the PARPM in terms of the parameters $p$ and $u$ 
(or $1/u$). The phases are: (a) non critical with no absorbing states; 
(b) critical with no absorbing states; (c) critical with infinitely many 
absorbing states (active); (d) inactive phase where the system is frozen 
in one of the infinitely many absorbing states. Line $u=0$ ($0\leq p <2$): the 
model has a gap (massive) but with a behavior that resembles a critical 
system (see section 3). 
Line $u=1$ ($0\leq p <8/3$): the model is critical and conformally invariant 
(see section 4). Line $1/u=0$ ($0\leq p <4$): the model is related to 
an extended TASEP model with non-local jumps of particles (see section 5).
The straight line $p=2$ ($0<u<\infty$) separates the 
phases (a) and (b), having no 
absorbing states, from the phase (c) with infinitely many absorbing states. 
The black line connecting $(u,p)=(0,2)$ to $(u,p)=(0,4)$ is a critical line 
separating the phase (c) with infinitely many absorbing states from the 
frozen phase (d).  
} 
\label{fig0}
\end{figure}

In order to check if this unusual behavior is linked to the conformal 
invariance of the model for $p<{p}_1$ we study the PARPM in regions where 
$u\neq 1$, where the model is either gaped ($u<1$),  
or  critical but not conformally 
invariant ($u>1$). 
 An overview of our results is given in the schematic phase diagram 
of the model shown in Fig.~\ref{fig0}.
In this paper we are going to restrict ourselves to the 
representative cases (red lines in Fig.~\ref{fig0}), where $u=1$, $u=0$ (no adsorption) and $u=\infty$ (no desorption), with 
arbitrary values of $p$.

The RPM although originally defined in an open chain can also be defined in 
a periodic lattice \cite{alcaraz2013nonlocal}. In the periodic chain the model can be 
interpreted as a particular extension of the asymmetric exclusion process 
(ASEP) where the particles (excluded volume) are allowed to perform local as 
well nonlocal jumps. We are going also to consider in this paper the PARPM 
formulated in periodic 
lattices. We verified that when $u\to \infty$ (only adsorption processes) 
the extended PARPM is exactly related to a totally asymmetric exclusion 
process (TASEP) 
where the particles jumps only in one direction. At $p=1$, where the model 
recovers the RPM, the model is mapped to the standard TASEP \cite{derrida1993exact,derrida1997exact}, and for 
$p\neq 1$ it can be interpreted as a TASEP whose transition rate to the 
neighboring sites depend on the total number of particle-vacancy pairs, 
in the configuration. 

At $u=0$ (no adsorption) the model is gapped but shows interesting properties. 
The configuration where there are no sites available for desorption is an 
absorbing state, since there is not adsorption process.  Although gapped the 
system stays during a large time, that increases polynomially with the 
lattice size, in a critical quasi-stationary state with dynamical critical exponent 
$z=1$. This phenomena is due to the appearance of an extensively large number of 
Jordan cells (or Jordan blocks), 
that for sufficiently large lattices produces a more important 
effect than the 
exponential decay induced by the gap. For some special initial conditions 
we are able to obtain analytically the Jordan cells and derive the time 
dependence of some of the observables. 
Actually Jordan cells are known to appear in models described by 
Temperley-Lieb algebra (TLA) and whose continuum limit are ruled by logarithmic 
conformal field theories \cite{pearce,morin}. Although the time evolution 
operator of the RPM is given by the sum of generators of the (TLA), 
the introduction of the parameter $p$, producing the PARPM, destroys 
this connection due to the appearance of nonlocal terms that are not 
expressed in a simple way in terms of the generators of the TLA.

The paper is organized as follows. In the next section we introduce the 
extension of the PARPM studied in this paper, for open and periodic boundary 
conditions. We present our results separately in sections 3, 4 and 5 for the 
model with $u=0$, $u=1$ and $u \to \infty$, respectively. Finally in section 6 we 
present our conclusions.

\section{ The peak adjusted raise and peel model} \label{sect2}

The raise and peel  model (RPM) and the peak adjusted model (PARPM) are models 
describing the stochastic time evolution of an interface defined by a set 
of integer heights  in a discrete lattice of size 
$L$ (even). These heights are $\{h_0,h_1,\ldots,h_L\}$ and  
$\{h_1,h_2,\ldots,h_L\}$ for the case of open and periodic boundary conditions, 
respectively. The height configurations satisfy the restricted solid-on-solid 
conditions 
\be \label{e2.1}
|h_{i+1}-h_{i}| =1, \quad h_i \in Z.
\ee
The boundary condition imposes the additional constraints: (a) for the 
periodic case 
$h_i=h_{L+1}$ ($h_i \in Z$) and  the configurations $\{h_i\}$ and 
$\{h_i+m\}$ ($m \in Z$)are the same, (b) for the open case  $h_0=h_L=0$ and $h_i$ are non-negative 
integers  ($h_i \geq 0$). 
 The  number of profile configurations of the surface is \cite{alcaraz2013nonlocal,de2004raise}:
\be \label{e2.2}
Z_b(L) = \frac{L!}{(\frac{L}{2})!(\frac{L}{2}+b)!},
\ee
where $b=0$ and $b=1$ for the periodic and open cases, respectively. 

\begin{figure}
\centering
\includegraphics[angle=0,width=0.5\textwidth] {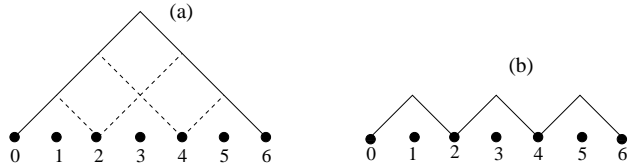}
\caption{
Special configuration of the PARPM with $L=6$ sites and open boundary 
conditions. a) The pyramid configuration. b) The substrate configuration.}
\label{fig1}
\end{figure}

The configurations have a local minimum (valley) or a local maximum (peak) 
whenever $h_{i-1}>h_i<h_{i+1}$ or 
 $h_{i-1}<h_i>h_{i+1}$, respectively.  We call for convenience, the 
configuration with the maximum number of peaks $N_{p}=L/2$ as
the substrate, and the configuration with the minimum number of peaks 
$N_{p}=1$ as pyramid. In Fig.~\ref{fig1} we show examples of these 
 configurations for the case where  the lattice has  $L=6$ sites and open 
boundary conditions.

In the dynamics of the RPM each site in a unit of time is visited with equal 
probability. The  PARPM however has the additional  parameter $p$  
that has the net effect of increasing ($p>1$) or decreasing ($p<1$)  the visiting 
probability for the sites with local peaks. For a given configuration the 
parameter $p \geq 0$ fix the probability that a particular peak is visited as 
$p_p=\frac{1}{L-b}p$, where $b=0,1$ for periodic and open ends. If a given 
configuration contains $N_p$ peaks, the probability $P_p$ that an arbitrary peak is 
visited is \cite{alcaraz2010conformal}
\be \label{e2.3}
P_p= N_p p_p = \frac{N_p p}{L-b},
\ee
while the probability that the remaining $N_{np}=L-b-N_p$ sites (sites with 
no peaks) are visited is given by
\be \label{e2.4}
P_{np}=1 - P_p = \frac{L-b-N_pp}{L-b}=q\frac{N_{np}}{L-b},
\ee
where the parameter $q$ gives the probability that a particular site, which 
is not a peak, is visited. This is not a new parameter, it depends on the 
parameter $p$ and the number $N_p$ of peaks in the configuration:
\be \label{e2.4p}
q=\frac{L-b-N_pp}{L-b-N_p}.
\ee
Since the maximum number of peaks is $\frac{L}{2}$,  $p$ should be restricted to the values:
\be \label{e2.5}
0\leq p \leq {p}_1,  \mbox{ where } {p}_1 = 2\frac{L-b}{L}, \quad (b=0,1).
\ee
 For $p=1$ all the sites are visited with equal probability independently 
of being peaks or not and we recover the standard RPM. For values of $p> 1$ 
($p<1$) the peaks have a larger (smaller) chance to be visited in a time 
interval.
At $p=0$ no peaks are visited in a unit of time. At $p={p}_1$ the 
visited sites in the configuration with $\frac{L}{2}$ peaks (substrate) are 
only the ones with peaks. 

 In this paper we generalize the definition of $p$ given in \rf{e2.3}, producing a richer model as we shall see. Given a configuration with $N_p$ peaks 
we now extend \rf{e2.3} by imposing that in a unit of time the peaks are 
visited with probability 
\ba \label{e2.6}
&&P_p = \mbox{Min} \left\{ 1,\frac{N_p}{L-b}p\right\}, \quad b=0,1,
\ea
where, as before $b=0$ ($b=1$) for periodic (open) boundary conditions. 
The probability $q$ that a particular site that is not a peak 
 is visited is given, using \rf{e2.6} and 
\rf{e2.4}, by 
\be \label{e2.6p}
q = \mbox{Max} \left\{ \frac{L-b-N_pp}{L-b-N_p},0\right\}.
\ee
Now 
 the parameter $p$ can be any non-negative number. For $p\geq {p_1}=2(L-b)/L$ all the 
configurations with number of peaks $N_p \geq \frac{L-b}{p}$ will be 
visited only at the sites containing peaks. 

We now define the dynamical rules for the stochastic evolution of the PARPM 
model. We can imagine a height configuration as a surface separating a solid 
phase from a gaseous one formed by tilted blocks that hit the surface. In a 
unit of time the blocks hit the sites with peaks ($h_{i-1}<h_i>h_{i+1}$), 
valleys ($h_{i-1}>h_i<h_{i+1}$), positive slope ($h_{i-1}<h_i<h_{i+1}$) or 
negative slope ($h_{i-1}>h_i>h_{i+1}$). The probability $P_p$ that we hit 
the sites that have a  peak is given by \rf{e2.6}, while the 
 probability that we reach the other non-peak sites is $P_{np} = 1-P_p$. In a unit of time the following processes may occur (see 
Fig.~\ref{fig2}), after a tile hits a site $i$:
\begin{figure}
\centering
\includegraphics[angle=0,width=0.65\textwidth] {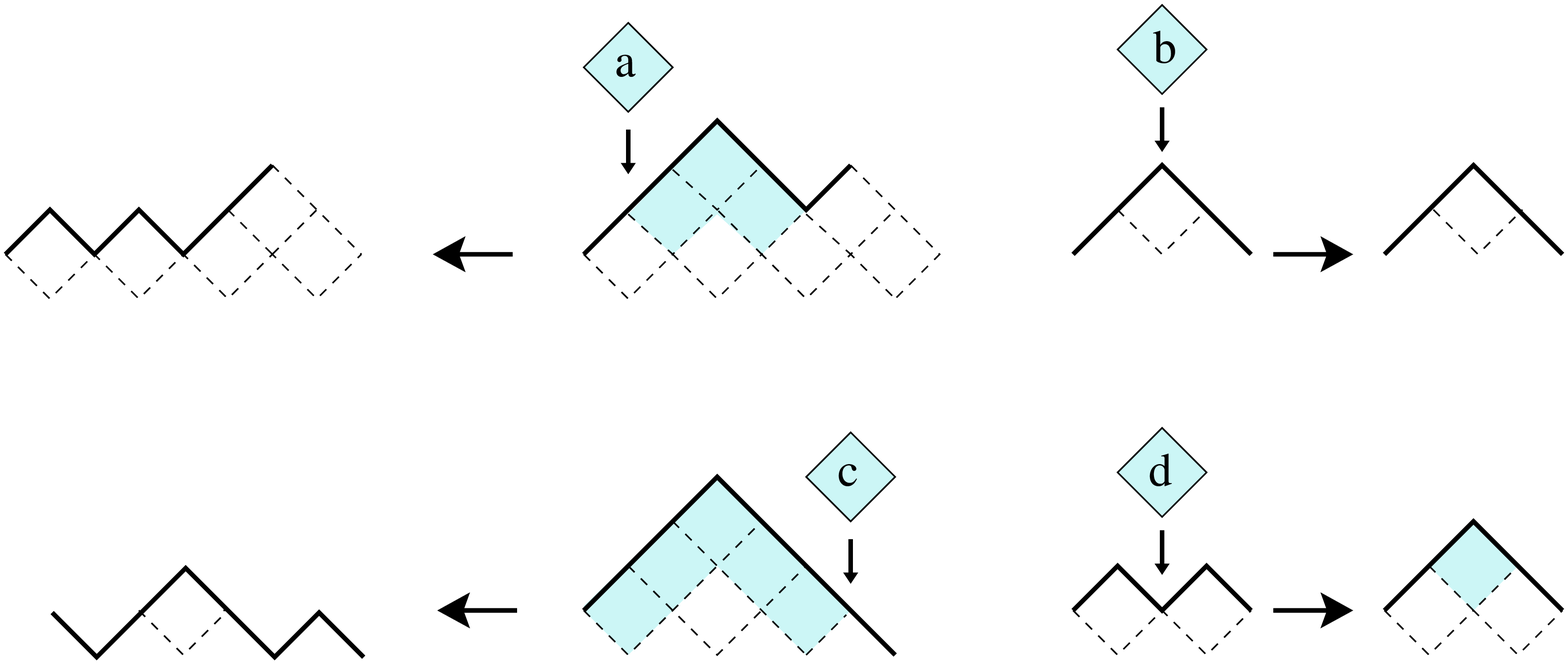}
\caption{
The dynamical processes in the PARPM. The tilted tile in the gaseous phase 
hit a positive slope (a), a peak in (b), a negative slope in (c) 
and a valley in (d) (see the text).}
\label{fig2}
\end{figure}

{\it{a}}) If we have a peak ($h_{i-1}<h_i>h_{i+1}$) the tile is reflected and 
the surface is unchanged (see Fig.~\ref{fig2}b).

{\it{b}}) If we have a valley ($h_{i-1}>h_i<h_{i+1}$), with probability 
proportional to the adsorption rate $u_a$, the configuration is changed 
$h_i \to h_i +2$ by adsorbing the tile 
(see Fig.~\ref{fig2}d).

{\it{c}}) If we have a positive slope ($h_{i-1}<h_i<h_{i+1}$), with a 
probability 
proportional to the desorption rate $u_d$, the tilted tile is reflected after 
 desorbing ($h_j \to h_j-2$) a layer of $t-1$ tiles from the sites $\{j=i+1,
\ldots,i+t-1\}$, where $h_j>h_i=h_{i+t}$ (see Fig.~\ref{fig2}a), i.e. 
 $t$ is  the minimum number of added sites to the right where the 
original height is repeated $h_i=h_{i+t}$.

{\it{d}}) If we have a negative slope ($h_{i-1}>h_i>h_{i+1}$), with a
 probability proportional to the desorption rate $u_d$, the tilted tile is 
reflected after desorbing ($h_j \to h_j-2$) a layer of $t-1$ tiles from the 
sites $\{j=i-t+1,\ldots,i-1\}$, where $h_j>h_i=h_{i-t}$ 
(see Fig.~\ref{fig2}c), i.e.
 $t$ is  the minimum number of subtracted  sites to the left where the    
original height is repeated $h_i=h_{i-t}$.

In a continuous time evolution, the probability $P_c(t)$ 
of finding the system in  the configuration $c=1,2,\ldots,Z_b(L)$ in a 
given time  
is given by the 
master equation: 
\be \label{e2.7}
\frac{dP_c(t)}{dt} = - \sum_{c'}H_{c,c'}P_{c'}(t),
\ee
that can be interpreted as an Schr\"odinger equation in imaginary time. The 
above defined stochastic rules give us the matrix elements $H_{c,c'}$ of the 
$Z_b(L)$-dimensional Hamiltonian. The Hamiltonian is an intensity matrix, 
i.e., the non-diagonal elements are negative and  $\sum_cH_{c,c'}=0$. 
This imply that the ground state of the system $\ket{0}$ has eigenvalue 0 
and its components give us the probabilities $P_c$ of the configurations 
in the stationary state, i.e., 
\be \nonumber
\ket{0}=\sum_cP_c\ket{c}, \quad P_c = \lim_{t  \to \infty}P_c(t).
\ee
The matrix elements $H_{c,c'}$ are calculated as follows. Only configurations 
that are connected through adsorption and desorption processes  give non-zero values. This means that 
apart from a multiplicative constant, that can be absorbed in the time scale, 
the matrix elements of the Hamiltonian is only a function of the ratio $u=\frac{u_a}{u_d}$. 
A configuration $\ket{c}$ with $N_p(c)$ peaks have $L-b-N_p(c)=N_{np}(c)$ 
sites where the desorption or adsorption processes may occur. Each of these 
$N_{np}(c)$ sites are reached with probability $q(c)$ defined in \rf{e2.6p}. 
 The non-diagonal elements $H_{c,c'}$ are given by $-u_aq(c')$, 
$-u_dq(c')$, or $-2u_dq(c')$, if the configuration $\ket{c}$ and $\ket{c'}$ 
are connected by a single adsorption,  one, or two distinct desorptions, 
respectively. 
The diagonal elements are calculated from the non-diagonal ones since 
$H_{c,c} = -\sum_{c \neq c'} H_{c,c'}$.

\begin{figure}
\centering
\includegraphics[angle=0,width=0.35\textwidth] {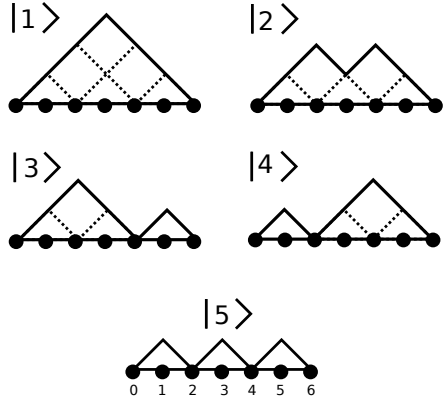}
\caption{
The five  configurations of the PARPM for a lattice with $L=6$ sites 
and open boundary conditions.}
\label{fig3}
\end{figure}

As an illustration let us consider the case of a lattice  size $L=6$ with 
open ends ($b=1$). In this case there are $Z_1(6)=5$ configurations 
$\ket{c}$ ($c=1,\ldots,5$), that are shown in Fig.~\ref{fig3}. The  number 
of peaks in each configuration give us, using \rf{e2.6p}: 
$q(1) = [(5-p)/4]_0$, $q(2)=q(3)=q(4)=[(5-2p)/3]_0$ and 
$q(5)=[(5-3p)/2]_0$, where we denote hereafter the maximum value:
\be \label{e2.8}
[x]_0 = \mbox{Max}\{0,x\}. 
\ee
The configuration $\ket{1}$ ($\ket{5}$) with lowest (largest) number of peaks 
is the one we called pyramid (substrate). The name substrate come from the 
fact that all the configurations can be obtained from it by the successive 
additions of tilted tiles. The configuration $\ket{1}$ is connected to 
$\ket{2}$ by a desorption at site 2 or 4. The configuration $\ket{2}$ is 
connected to the configuration $\ket{1}$ by an adsorption at site 3 and to the 
configurations $\ket{4}$ and $\ket{3}$ by a desorption at site 1 and 5, 
respectively. Following similarly with the other configurations we obtain the 
$5\times5$ Hamiltonian:
\ba \label{e2.9}
 H =-
\left( \begin{array}{c|rrrrr}
 & \bra{1} & \bra{2} & \bra{3} & \bra{4} &\bra{5} \\ \hline
\ket{1} &   [4u_d\frac{5-p}{4}]_0 & -[u_a\frac{5-2p}{3}]_0 &  0   &  0 & 0 \\
\ket{2} &   -[2u_d\frac{5-p}{4}]_0  & [(u_a+2u_d)\frac{5-2p}{3}]_0  &  
-[u_a\frac{5-2p}{3}]_0  & -[u_a\frac{5-2p}{3}]_0 &      0 \\
\ket{3} &  0   & -[u_d\frac{5-2p}{3}]_0     &  [(u_a+2u_d)\frac{5-2p}{3}]_0 &0 & -[u_a\frac{5-3p}{2}]_0 \\
\ket{4} &  0   & -[u_d\frac{5-2p}{3}]_0     & 0 &[(u_a+2u_d)\frac{5-2p}{3}]_0 & -[u_a\frac{5-3p}{2}]_0 \\
\ket{5} & -[2u_d\frac{5-p}{4}]_0     &  0 & -[2u_d\frac{5-2p}{3}]_0  &
-[2u_d\frac{5-2p}{3}]_0 &   [2u_a\frac{5-3p}{2}]_0 
\end{array} \right) .&&\nonumber \\
\ea
We observe that for arbitrary values of $p\geq0$, when the adsorption 
(desorption) is absent $u_a=0$ ($u_d=0$) the matrix elements connecting the 
configuration $\ket{5}$ ($\ket{1}$) that we called substrate (pyramid) are 
zero. The configurations $\ket{1}$ and $\ket{5}$ are in this case absorbing 
states. This is valid for general lattice sizes. Once the system reaches these 
configurations it stays on them forever. The Hamiltonian \rf{e2.9} for 
$0\leq p\leq {p}_1=5/3$ and $u=1$ recovers the one in the original 
formulation of the PARPM, introduced in \cite{alcaraz2010conformal}. In the extended version of 
the model \rf{e2.9} we see  that  for arbitrary values of $u$ we have no absorbing states 
as long  $0\leq p<{p}_1 =5/3$. At $p={p}_1=5/3$ the configuration $\ket{5}$ 
(substrate) becomes an absorbing state. For 
${p}_1 \leq p<{p}_2=5/2$, the configuration $\ket{5}$ is the single absorbing state. For ${p}_2\leq p <{p}_3=5$, the configurations $\ket{2}$, 
$\ket{3}$ and $\ket{4}$ containing two peaks also become absorbing states. 
Finally for $p\geq {p}_3$ all the configurations are absorbing states. 

The proliferation of absorbing states when $p \geq {p}_1$  is one of the main features of the 
PARPM. For arbitrary values of $u=u_a/u_d$ and lattice sizes $L$, it is 
convenient to  define
\be \label{e2.9p}
{p}_i=2\frac{L-b}{L+2-2i}, \quad i=1,2,\ldots,L/2, 
\ee
where $b=0,1$ for periodic and open boundaries. For 
${p}_i\leq p<{p}_{i+1}$ all the configurations with 
$N_p\geq \frac{L}{2}-i+1$ are absorbing states. The number of configurations with 
$N_p=\frac{L}{2}-i+1$, grow very fast as $i$ increases and we should expect 
an interesting behavior once $p>{p}_1$.

The model at $u=1$ and for $p< {p}_1$ is known to be critical and 
conformally invariant \cite{alcaraz2010conformal}. At $p=1$, where the model recovers the 
standard RPM, the model in non-critical (gapped) for $u<1$ and gapless for 
$u\geq 1$. For an overview of the phase diagram of the model see 
Fig.~\ref{fig0}. In the next sections we are going to study the model for arbitrary 
values of $p$, but considering only the three representative cases where 
$u=0$, $u=1$ and $u=\infty$ (red lines in Fig.~\ref{fig0}), 
that we  have distinct physical 
properties. In these sections we are going to measure several 
observables. The average height at time $t$ is defined as:
\be \label{e2.10}
h_L(t) = \frac{1}{L} \sum_{i=1}^L <h_i(t)>,
\ee
where $<h_i(t)>$ is the average height at the site $i$. In a given 
configuration, we define as contact points the points where the profile touch 
the substrate. In the open boundaries case their average are: 
\be \label {e2.11}
K_L(t)= <\sum_{i=1}^{L-1} \delta_{h_i(t),0}>,
\ee
while for the periodic case:
\be \label{e2.12}
K_L(t) = <\sum_{i=1}^{L} \delta_{h_i(t)-h_{\scp{\mbox{min}}},0}>,
\ee
where $h_{\mbox{\scp{min}}}$ is the minimum height in the configuration. 
Another interesting quantity is the number of peaks and valleys in the 
configuration.  The related observable is:
\be \label{e2.13}
\tau_L(t) = \frac{1}{L-b} \sum_{i=1}^{L-b} \left(1-\frac{|h_{i-1}-h_{i+1}|}{2}
\right), 
\quad b=0,1.
\ee 

\section{The model with no adsorptions}

We consider in this section the limiting case of the PARPM where we have 
no adsorptions, i. e., $u=u_a/u_d=0$. 
In this case by ordering the configurations according to their 
number or peaks, i. e.,  the  first state is the pyramid 
($N_p=1$) and the last one the substrate ($N_p=L/2$), the Hamiltonian in the 
master equation \rf{e2.7} is a lower triangular matrix. This happens because 
we have only desorptions and their effect in the configurations is the 
increase of the number of peaks. An example for $L=6$ with open boundary 
condition ($b=1$) is obtained    by  setting  $u_a=0$ in 
\rf{e2.9}. Since $H$ is a lower triangular 
matrix its $Z_b(L)$ eigenvalues are given by its diagonal elements $H_{c,c}$ 
($c=1,\ldots,Z_b(L)$). The diagonal elements in the PARPM are given by 
\be \label{e3.1}
H_{c,c} = (N_s^{(c)} u_d + N_v^{(c)} u_a)q(c),
\ee
where $N_s^{(c)}$ and $N_v^{(c)}$ are the number of sites where we have a nonzero 
slope and valleys, respectively. The $p$-dependent parameter $q(c)$ is given 
by \rf{e2.6p}. For the case of no adsorption, we then have the 
eigenvalues
\be \label{e3.2}
E(N_p,p,L,u_d) = \mbox{Max} \left\{0, u_d(L-2N_p)\frac{L-b-N_pp}{L-b-N_p}\right\},
\ee
where we have used \rf{e2.6p} and the fact that $N_s=L-2N_p$. 

The present case, where $u_a=0$, the results for the periodic case can be 
obtained from the open  
 boundary case 
straightforwardly. We are going, in the rest of this section, 
to restrict ourselves 
to the open boundary case where $b=1$.
It is convenient to label the energies \rf{e3.2} by the index $i=L/2 -N_p$: 
\be \label{e3.3}
E_i=\mbox{Max} \left\{0,2i u_d\frac{(2-p)L +2(ip-1)}{L+2(i-1)}\right\}, \quad 
i=0,1,\ldots,L/2
\ee

\begin{figure}
\centering
\includegraphics[angle=0,width=0.35\textwidth] {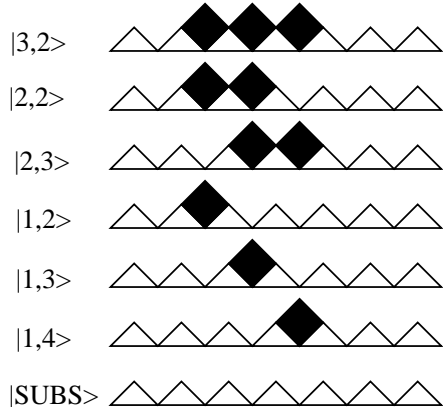}
\caption{
Configurations $\ket{N_0,k}$ with $(\frac{L}{2}-1)$ peaks obtained  by 
adding, in the substrate configuration, $N_0$ tiles at the closest positions. The 
index $k$ indicate the valley position in the substrate where the initial 
tile is added. The substrate configuration $\ket{SUBS}$ is also shown. }
\label{fig4}
\end{figure}

Let us restrict ourselves initially to the cases where the parameter $p\leq 2$.  The stationary state of the system is the substrate, being  the 
ground-state ($E_0=0$) of the Hamiltonian. In this no adsorption case 
the ground state  is 
also an absorbing state. If the initial configuration is not the substrate, 
for sufficiently large times the dynamics will be ruled by the first gap of 
the Hamiltonian $E_1-E_0=E_1$, given by \rf{e3.3}, and since $E_1$ is finite 
for any $L$, we do expect for any observable an exponential time-decay 
$\tau\sim1/E_1$. However we need some care in this analysis since the 
eigenvalue $E_1$ has a degeneracy  
$d_{\frac{L}{2}-1}=(\frac{L}{2}-1)(\frac{L}{4})$. This is the number of configurations with $N_p=\frac{L}{2}-1$ 
peaks, obtained by adding tiles in the first layer ($h_i\leq 2$) above the 
substrate and with the tiles in the  closest positions (no nonzero slopes between 
the added tiles) (see Fig.~\ref{fig4} for examples). For large $L$ the 
degeneracy grows as $L^2$ and as we shall see, an interesting physical 
behavior occurs.

To simplify our analysis we are going to consider as the initial state, for 
the evolution of our 
system, one of the configurations with $L/2-1$ peaks, and  with $N_0$ tiles 
added in the substrate. There are $L/2-N_0$ configurations of this kind and 
we denote them as $\ket{N_0,k}$, $k=1,\ldots, L/2-N_0$; $N_0=1,\ldots,L/2-1$. 
As an example, we show in Fig.~\ref{fig4} some of the configurations for a 
lattice size $L=14$, together with  the substrate absorbing state 
$\ket{\mbox{SUBS}}$. 

If we chose for $L=14$ the state $\ket{3,2}$ (see Fig.~\ref{fig4}) as the 
initial state, the effective Hamiltonian is given by
\ba \label{e3.4}
&& H =-
\left( \begin{array}{c|rrrrrrr}
 & \ket{3,2} & \ket{2,2} & \ket{2,3} & \ket{1,2} &\ket{1,3} & \ket{1,4} & 
\ket{\mbox{SUBS}}\\ \hline
\bra{3,2} &  E_1 & 0 & 0 & 0 & 0 & 0 & 0  \\
\bra{2,2} &   -E_1/2  & E_1  &  0  & 0 &      0 & 0& 0 \\
\bra{2,3} &  -E_1/2 & 0 & E_1 & 0 & 0 & 0 & 0 \\
\bra{1,2} &  0   & -E_1/2 & 0& E_1 & 0 &0 & 0 \\
\bra{1,3} &  0   &-E_1/2 &  -E_1/2 & 0 & E_1 & 0 & 0 \\
\bra{1,4} &  0   &0 &  -E_1/2 & 0 & 0 &  E_1  & 0 \\
\bra{\mbox{SUBS}} & 0 & 0 & 0 & -E_1 & -E_1 & -E_1 & 0 \\
\end{array} \right) ,
\ea
where $E_1$ is given by \rf{e3.3}. It is interesting to note that $H$ in 
\rf{e3.4} has a 
Jordan-cell structure, since  although the eigenvalue $E_1$ is 6 degenerated 
there are only 3 distinct eigenvectors. In the general case, for arbitrary $L$, 
if we start with an configuration with $L/2-1$ peaks and $N_0$ tiles added 
in the substrate, we have a Jordan-cell structure where instead of 
$N_0(N_0+1)/2$ eigenvectors with eigenvalue $E_1$, we have only $N_0$ distinct 
eigenvectors ($N_0(N_0-1)/2$ are missing!). If we start, for large $L$, with 
an initial state where $N_0\sim L$, the Jordan-cell will be of dimension 
$\sim L^2$.

Let us consider, as an example, the  probability distribution $\{\ket{P(t)}\}$ 
for the configurations of the 
$L=14$ system considered in Fig.~\ref{fig4} when we take as initial state 
the configuration $\ket{3,2}$.   Solving the set of coupled 
linear differential equations derived by inserting the effective Hamiltonian 
\rf{e3.4} in the master equation \rf{e2.7} we obtain:   
\ba \label{e3.5}
\ket{P(t)} &=& (1-\sum_{k=0}^2 \frac{(E_1t)^k}{k!} e^{-E_1t}) \ket{\mbox{SUBS}}
 + \frac{e^{-E_1t}}{2!} \ket{3,2} 
+(E_1t)e^{-E_1t}\frac{\ket{2,2}+\ket{2,3}}{2} \nonumber \\
&&+\frac{(E_1t)^2}{2!}e^{-E_1t}\frac{\ket{1,2}+2\ket{1,3}+\ket{1,4}}{2^2}.
\ea
We see that the missing 3 eigenvectors of the Jordan-cell structure produces 
besides the exponential decay also a polynomial  time dependence. The 
configuration with $N_0- k$ tiles above the substrate has a time dependence 
$t^ke^{-E_1t}$.

We can generalize the above solution for $L$ arbitrary. Let us 
consider, as the initial 
state, one of the configurations $\ket{N_0,k}$ (see Fig.~\ref{fig4}) with $\frac{L}{2}-1$ peaks and $N_0$ tiles 
added in the first layer above the substrate. The probability of  finding the 
system at time $t$ in a configuration with $(N_0-k)$ tiles is given by:
\be \label{e3.6}
P_k(t) = \frac{(E_1t)^k}{k!} e^{-E_1t}, \quad 1\leq k \leq N_0-1,
\ee
while the probability of  finding  the system in the substrate configuration is 
given by: 
\be \label{e3.7}
P_{\mbox{\scp{SUBS}}} = 1-\sum_{k=1}^{N_o-1} P_k(t).
\ee
\begin{figure}
\centering
\includegraphics[angle=0,width=0.4\textwidth] {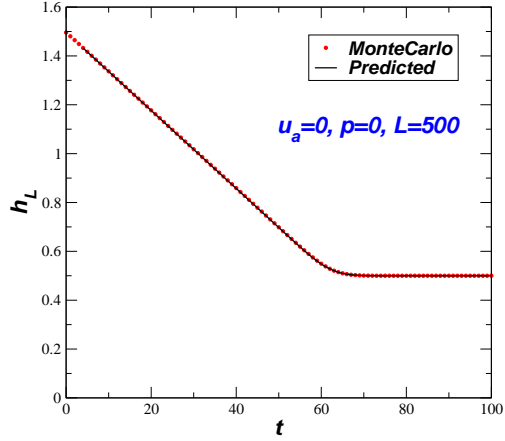}
\caption{
Average height $h_L(t)$ as a function of time for the PARPM with $u_a=0$, 
$p=0$ and lattice size $L=500$. The initial configuration is one of the 
$\ket{N_0,k}$, with $N_0=249$ (see  Fig.~\ref{fig4} for notation). The continuous curve is the 
prediction given by \rf{e3.10}, and the discrete points are the results 
obtained from Monte Carlo simulations.}
\label{fig5}
\end{figure}
\begin{figure}
\centering
\includegraphics[angle=0,width=0.4\textwidth] {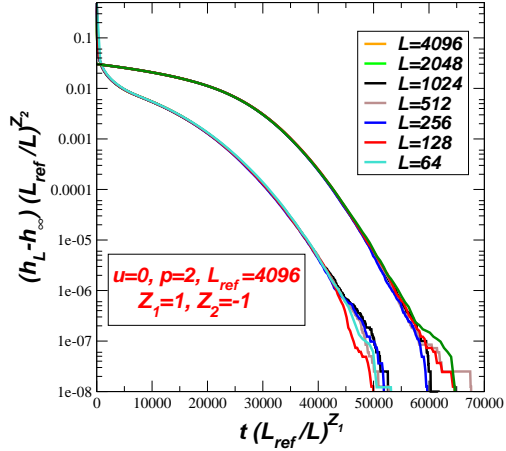}
\caption{
The finite-size scaling of the average height  for the PARPM with $u=0$ and $p=p_c=2$. The two 
set of curves correspond to distinct initial conditions. In the  lower set the 
initial configuration is  $\ket{N_0,1}$ (see Fig.~\ref{fig5}) 
where we have $N_0=31$ tiles in the first row above the substrate. In the 
top curves the initial configuration is the one where  we have a pyramid 
 whose basis has $N_0=31$ in the first row, above  the substrate configuration.
The parameter $L_{ref}=4096$ was chosen to better present the results and 
the exponents are $z_1=z_2=1$. }
\label{fig6}
\end{figure}

From \rf{e3.6} and \rf{e3.7} we can calculate the time dependence of the 
observables \rf{e2.10}-\rf{e2.13}. The average number of peaks and valleys 
$\tau_L(t)$, heights $h_L(t)$ and contact points $K_L(t)$ are given by
\be \label{e3.8}
\tau_L(t)-\tau_L(\infty) = -\frac{2}{L-1}\sum_{k=0}^{N_0-1} 
\frac{(E_1 t)^k}{k!} e^{-E_1t},
\ee
\be \label{e3.9}
(h_L(t)-h_L(\infty))\frac{L}{2}=K_L(\infty)-K_L(t) = 
e^{-E_1t}\left\{\sum_{k=0}^{N_0-1}(N_0-k)\frac{(E_1t)^k}{k!}\right\},
\ee
where $\tau_L(\infty)=1-\frac{1}{L}$, $h_L(\infty)=\frac{1}{2}$ and 
$K_L(\infty) = \frac{1}{2}-\frac{1}{L}$ are the values of these quantities 
at the final substrate configuration.
In both expressions, if $(E_1t)^{N_0-1}/(N_0-1)! > (E_1t)^{N_0}/N_0!$, i.e., 
$t<t_0=N_0/E_1$, we can approximate the sum by 
$\exp(E_1t) + O((E_1t)^{N_0}/N_0)$. This is quite interesting since for times 
up to $t\lesssim t_0$ the exponential decay is totally canceled and the system 
behaves as if it was  gapless. Only for $t\gtrsim t_0$ the exponential decay will 
govern the time decay to the stationary state. In Fig.~\ref{fig5} we 
illustrate this behavior by comparing the average height $h_L(t)$ 
predicted by \rf{e3.9} with the results obtained by averaging $3200$  
runs of a Monte Carlo simulation for the PARPM with $L=500$ (open ends), with 
parameters $u_d=1$, $p=0$ and having as the initial state  the configuration 
containing $N_0=249$ tiles in the 
first layer above the substrate. The agreement is excellent, as expected. 
We also see in the 
figure that up to a transient time $t_0=N_0/E_1$ ($\sim62$ in this case), 
 $h_L(t)$ has a linear time dependence, and for $t\gtrsim t_0$ the system 
finally decays to the substrate, where $h_L(\infty)=1/2$. This linear behavior 
is better seen by writing \rf{e3.9} as 
\be \label{e3.10}
<h_L(t)-h_L(\infty)>\frac{L}{2} = e^{-E_1t}\left\{ \frac{N_0}{(N_0-1)!}
(E_1t)^{N_0-1} + (N_0-E_1t)\sum_{n=0}^{N_0-2}\frac{(E_1t)^n}{n!}\right\},
\ee
where the linear behavior dominates up to $t\lesssim t_0=N_0/E_1$.

Let us analyze the time behavior for large lattices and $0<{p}<p_1=2$. 
From \rf{e3.3} the gap is finite and given by 
\be \label{e3.11}
E_1=2u_p(2-p +2(p-1)/L).
\ee
In order to do the limit $L\to \infty$ is necessary to  define what are the initial 
states, as we change the lattice sizes. Two distinct limits give us quite 
different behavior. 
In the first one we consider a set of initial states where we have a fixed 
number of tiles $N_0$ above the substrate. In this case we have the linear 
time behavior up to the finite time $t_0=N_0/E_1$, but for $t\gtrsim t_0$ the 
system exhibits its exponential-law decay. In the second case we consider the 
set of initial states that $N_0/L$ is fixed as $L\to\infty$. In this case 
$t_0=N_0/E_1 \to \infty$, and the system behaves at all times as would not be 
gapped, similarly as happens in a critical system. 
This unusual and surprisingly behavior 
happens due to the infinite size Jordan-cell structure for the eigenvalue 
$E_1$ of the Hamiltonian.

Let us now consider  the case $p={p_1}=2$. In this case we see from 
\rf{e3.11} that the gap vanishes as $E_1\sim 1/L$, when $L\to \infty$, and 
the system is critical. In order to see this critical behavior we must 
consider, as $L\to \infty$, a sequence of initial states with  a fixed number $N_0$ of tiles above the substrate. In Fig.~\ref{fig6} we illustrate this critical 
behavior by considering, for several lattice sizes the average height $h_L(t)$ 
for two distinct sets of initial conditions. In the first one we take as the 
initial configurations the configurations $\ket{N_0,1}$ given in 
Fig.~\ref{fig5}, where we have $N_0=31$ tiles added in the first row above 
the substrate, while in the second one we take as initial configurations the ones 
where we have a pyramid whose basis contains  $N_0=31$ tiles in the first row above 
the substrate.  As we can see in Fig.~\ref{fig6} for both initial conditions, the 
curves for the several lattice sizes collapse 
 in the finite-size scaling curve 
\be \label{e3.12}
\frac{h_L(t)}{h_{\infty}}-1 = \frac{1}{L^{\alpha}} f(\frac{t}{L^z}), 
\ee
with the dynamical critical exponents $\alpha=z=1$. For $p\geq 2$, the initial states 
we considered are already absorbing states and the system is already in the 
stationary state. It is simple to convince ourselves that even if we consider 
more general initial states, that are not absorbing states, since we have 
only desorption, after a relatively short time the system is trapped on one 
of the absorbing states. This means that for $u=u_a/u_d=0$ we have a phase 
transition in the PARPM at $p=p_c=2$, separating  phases with  single 
and multiple absorbing states. 
\cite{hinrichsen2000non}. 
Knowing that $z=1$, we can write from \rf{e3.11} the gap $E_1 \sim L^{-z}g(\Delta^{1/\nu_{\perp}})$, with $\Delta=p-p_c$, $z=\nu_{\perp}=1$, that 
is similar to the results obtained from the numerical diagonalization of the 
evolution operator of the contact process.

It is important to mention that although $z=1$ the model at $p=p_c=2$ is not 
conformally invariant. From \rf{e3.3} we see that, for $L\to \infty$, the gaps 
behave as $E_i \sim A_i/L$ with amplitudes $A_i=4iu_d(2i-1)$. 
These amplitudes does not give the conformal towers that are expected in the 
finite-size scaling limit of conformally invariant critical systems.

\section{The model with equal rates of adsorption and \\desorption}

We consider in this section the PARPM for the special case where the 
absorption and desorption rates are equal, i. e, $u=u_a/u_d=1$. For this case  
the model with open boundary condition was studied in \cite{alcaraz2011conformal} for the 
values of the parameter $0\leq p<{p}_1=2-\frac{1}{L}$. The model is 
conformally invariant for $0\leq p <{p}_1$, with a $p$-dependent sound 
velocity. At $p={p}_1$ the substrate configuration becomes an absorbing 
state and is the stationary state. However, the system stays during 
a time interval which  grows exponentially with the lattice size, in a 
quasi-stationary state with the same properties ($z=1$) as the conformal 
invariant phase $0<p<{p}_1$ of the  model.
\begin{figure}
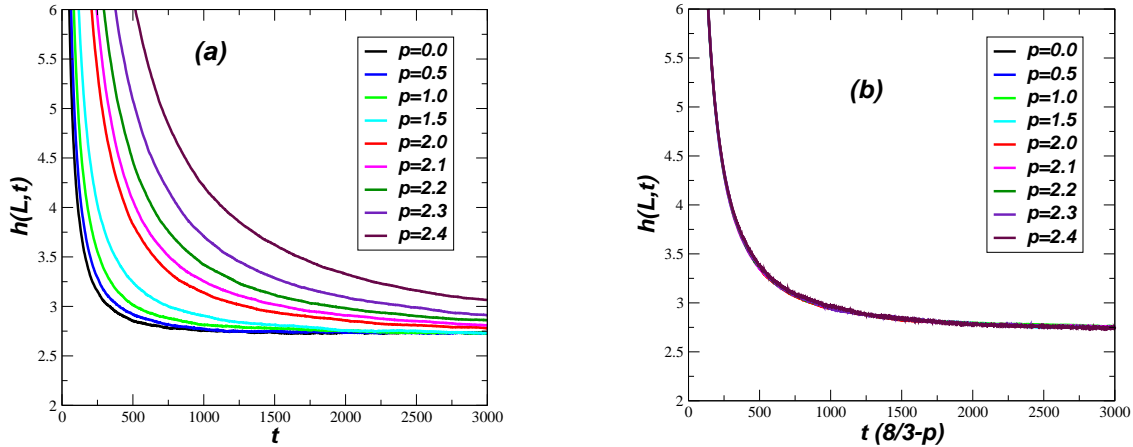

\begin{subfigure}{.5\textwidth}
\centering
\includegraphics[angle=0,width=0.8\textwidth] {fig7a-diego.eps}
\label{fig7a}
\end{subfigure}
\begin{subfigure}{.5\textwidth}
\centering
\includegraphics[angle=0,width=0.8\textwidth] {fig7b-diego.eps}
\label{fig7b}
\end{subfigure}
\caption{(a) Values of the average height $h_L(t)$ for the 
PARPM with $L=8192$ sites and open boundary conditions. The curves are 
for several values of the parameter $p$ and $u=1$.  (b) 
Same as (a) but the time is now scaled by the factor $(8/3-p)$, which is 
proportional to the sound velocity.} 
\label{fig7}
\end{figure}

\begin{figure}
\centering
\includegraphics[angle=0,width=0.4\textwidth] {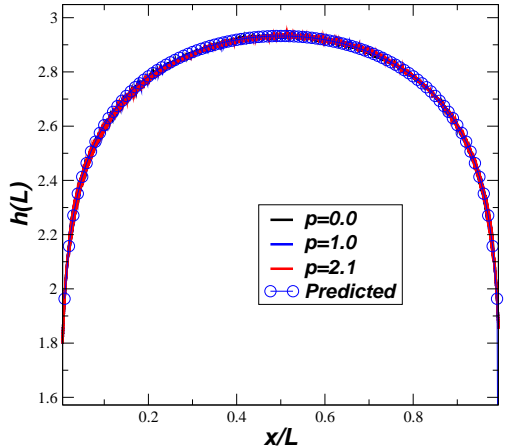}
\caption{
The average distribution of heights at the sites $x$ ($0 \leq x \leq L$)   in the stationary state 
of the PARPM with open boundary conditions. The lattice size is  $L=8192$, $u=1$ and 
the parameters $p$ are shown in the figure. The predicted universal curve $h(L) = 0.2757\ln[L\sin(\pi x/L)/\pi] + 0.77$ is also shown (circles).} 
\label{fig8}
\end{figure}

We now consider the case where $p>{p}_1$. In Fig.~\ref{fig7}a we show the 
time evolution of the average height $h_L(t)$ for the model with open boundary 
conditions, lattice size $L=8192$ and some values of $p$. In Fig.~\ref{fig7}b  we show that 
these  curves, similarly as happens for $p<{p}_1$ \cite{alcaraz2010conformal}, are 
collapsed in a single universal curve by re-scaling the time scale by a factor 
proportional to the sound velocity of the model $t \to t(\frac{8}{3}-p)$. 
Similar curves are also obtained for the other observables defined in 
\rf{e2.12}-\rf{e2.13}. This is a surprise, since for $p>{p}_1$, as $p$ 
increase the number of absorbing states increases drastically. For 
$L\to \infty$, although the system has an infinite number of absorbing states, 
it stays in a quasi-stationary state with similar properties as the 
conformally invariant systems ($0\leq p<{p}_1$). 
The system prefers (independent of the value of $p$) to stay in configurations 
with number of peaks smaller than those of the absorbing states, and the 
net effect of the parameter $p$ is just a change in the time scale. 
To illustrate, we show in
Fig.~\ref{fig8} the average distribution of heights $h_L(x)$ at site $x$ for 
the model in the quasi-stationary state. The data are for $L=8192$ sites, open 
boundary conditions and the values of $p=0,1$ and $2.1$. We see a nice 
collapse of the curves regardless $p<{p}_1$ or $p>{p}_1$. 
We also show in the figure (circles) the predicted curve \cite{ARS} for the
standard $u=1$ RPM: 
\ba  h(L) = \gamma\ln[L\sin(\pi x/L)/\pi] + \beta , \nonumber 
\ea
with
$\gamma=\sqrt{3}/2\pi \approx 0.2757$ and $\beta=0.77$. The coefficient
$\gamma$ can be derived exploiting the conformal invariance of the model
\cite{JS,ARS}.
We 
verified that the curves collapse up to $p=p_c=8/3$. At this point, 
from \rf{e2.9p}, all the configurations with $3L/8$ peaks are absorbing 
states, and the time scale 
disappears (the sound velocity is zero). For $p>p_c$ the system decays, after 
a short time, in one of the infinitely many absorbing states. 

\begin{figure}
\centering
\includegraphics[angle=0,width=0.4\textwidth] {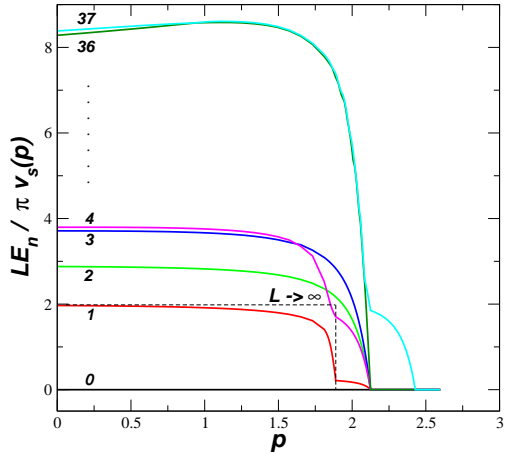}
\caption{
The ground state ($E_0$) and the lowest 37 excited eigenenergies, 
as a function of $p$,  of 
the Hamiltonian of the PARPM, with 
$u=1$, lattice size $L=18$ and open boundary conditions. In the figure
 $v_s(p)=9(8/3-p)\sqrt{3}/10$ is the sound velocity. The dashed line 
 is the prediction for the first excited state for $p<8/3$ when 
$L\to \infty$.} 
\label{fig9}
\end{figure}

In order to calculate the decay time of the quasi-stationary state for 
${p}_1 <p \leq p_c$, we need to evaluate the  eigenvalues of the 
Hamiltonian. In Fig.~\ref{fig9} we show the lowest 38 eigenvalues of the 
Hamiltonian as a function of $p$, for the model with $L=18$ sites and open 
boundary conditions. We see from this figure that at $p={p}_1=
\frac{17}{9}=1.888\cdots$  the energy of the eigenstate  that is 
 the first excited  state 
for $p<{p}_1$ has a large decrease. It degenerates with the ground 
state ($E_0=0$)   
exponentially with the lattice size, as $L \to \infty$. This 
is due to the appearance of the absorbing state whose configuration is the 
substrate, with 9 peaks. At $p={p}_2=\frac{17}{8}=2.125$ additional 
36 excited states also degenerate with the ground state, since all the 
configurations with 8 peaks are also absorbing eigenstates. At $p={p}_3=
\frac{17}{7}\approx 2.428$ additional degeneracies of the ground 
state happens 
due to the new absorbing states with 7 peaks. We also verify in 
Fig.~\ref{fig9} that the lowest gap is much lower in the
  region $p>{p}_1$ 
as compared with the one in the conformally invariant region $p<{p}_1$. In the region $p<{p}_1$ we verified from 
a finite-size extrapolation, using lattice sizes up to 
$L=28$, that the first excited state, 
as $L\to \infty$, is 
given by $E_1=E_1(p)=2\pi v_s(p)/L$ where 
$v_s(p)= 9(\frac{8}{3}-p)\sqrt{3}/10$ is the sound velocity, 
in agreement with 
the prediction in Ref.~\cite{alcaraz2010conformal}. 
This value of $E_1$ for  $L\to \infty$  is represented as the dashed line 
in Fig.~\ref{fig9}. The fact that even for $p>{p}_1$ the system has 
similar physical properties as $p<{p}_1$ is an indication that most 
probably the lowest gap above the quasi-stationary state is also given by 
the same  last expression $E_1(p)=2\pi v_s(p)/L$. 

For $p\geq{p}_1$ it is not possible to obtain reasonable estimates for 
the lowest gap $E_1$, from direct diagonalizations of the Hamiltonian. In this
case the finite-size effects are stronger and we need to consider 
lattices sizes larger than the ones we were able to diagonalize exactly 
($L\sim 20$). In this case we extract $E_1$ from the time evolution of the 
average number of peaks and valleys 
\be \label{e4.1}
\tau_L(t)=\tau_L(\infty)[1-a\exp(-E_1t)],
\ee
where $a$ is a constant that depends on the initial condition considered. 
In order to test the precision of the estimates obtained from \rf{e4.1} we 
compare them with the exact results derived from  direct diagonalizations 
of the Hamiltonian with  small lattice sizes and $p={p}_1=2$. In table 1 we give the 
estimates obtained from both methods. As we can see for $L>8$ the difference 
among the estimated values are less that $1\%$. 

In order to estimate 
$E_1$ using larger lattice sizes we chose, for a fixed value of $p=p_{i+1}$, a 
sequence of lattice sizes $L=L(i)$ such that, from \rf{e2.9p},  $p=(L-b)/(\frac{L}{2}-i)$, with 
$i$ an integer and $b=0$ or 1 for the periodic or open boundary conditions. 
For example for $p=2.2$ the lattice sizes are $L(i)=22i$ and 
$L(i)= 12 +22(i-1)$ for periodic and open boundary conditions. 
\begin{table*}[htp]
\label{tab1}
\begin{center}
\begin{tabular}{lcccccr}
\cline{1-7}

   $E_1(L)$        & 8 & 10 & 12 & 14 & 16 & 18 \\ \hline 
  Exact & 0.227 & 0.139 & 0.0922 & 0.0636 & 0.04481 & 0.0318 \\
  From  \rf{e4.1} & 0.231 & 0.140 & 0.0920 & 0.0638 & 0.04476 & 0.0320 \\
\hline
\hline
\end{tabular}
\end{center}
\caption{ Estimated results for the lowest gap $E_1$ for the PARPM with 
$u=1$ and $p=2$ and lattice sizes $L=8-18$. In the first row we show the 
 values obtained from the 
exact diagonalization of the Hamiltonian. In the second row the values 
 are obtained from the large-time fit (using \rf{e4.1}) of $5\times 10^5$ samples in  
a Monte Carlo simulation.}
\end{table*}

\begin{figure}
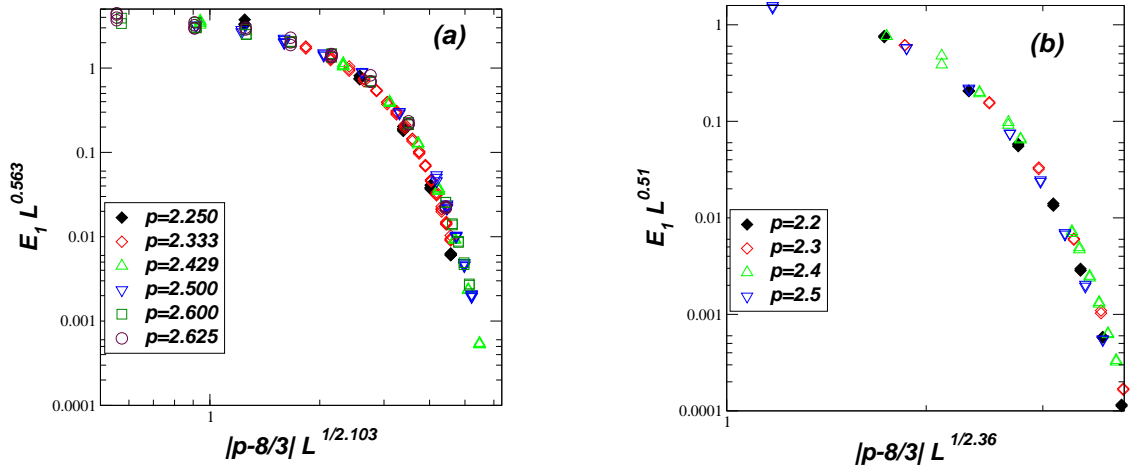

\begin{subfigure}{.5\textwidth}
\centering
\includegraphics[angle=0,width=0.8\textwidth] {fig10a-diego.eps}
\label{fig10a}
\end{subfigure}
\begin{subfigure}{.5\textwidth}
\centering
\includegraphics[angle=0,width=0.8\textwidth] {fig10b-diego.eps}
\label{fig10b}
\end{subfigure}
\caption{Finite-size scaling for the gap of the PARPM with $u=1$, 
for several values of $p$  
 and lattice sizes 
$L \leq 18292$ (compare with \rf{e4.1}). The lattice size values   depend on the  parameter 
$p$ (see the text). 
(a) Open 
boundary condition. b) Periodic boundary condition.}
\label{fig10}
\end{figure}

\begin{figure}
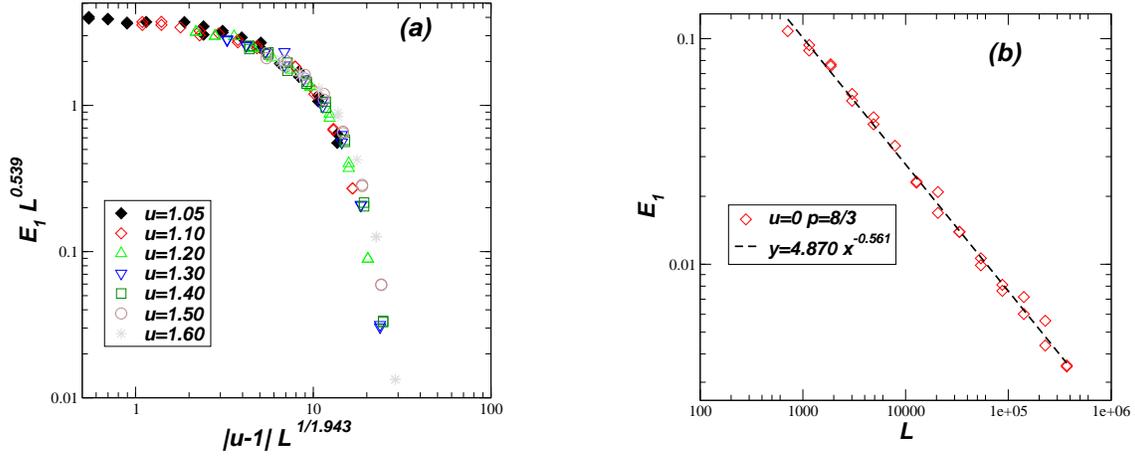

\begin{subfigure}{.5\textwidth}
\centering
\includegraphics[angle=0,width=0.8\textwidth] {fig11a-diego.eps}
\label{fig11a}
\end{subfigure}
\begin{subfigure}{.5\textwidth}
\centering
\includegraphics[angle=0,width=0.8\textwidth] {fig11b-diego.eps}
\label{fig11b}
\end{subfigure}
\caption{Energy gap $E_1$, as a function of $L$,  for the PARPM with periodic boundary 
conditions for the parameter $p=8/3$ and in the region 
$u\geq 1$.  (a)  Data collapse using in \rf{e4.2} $\Delta = u-u_c$, 
with $u_c=1$. The lattice sizes are $104\leq L \leq 54120$. (b)  
Power-law behavior for the gap at $u=1$. The lattice sizes 
are $L=8.i$ ($712\leq L \leq 370944$), so that $p=L/(L/2-i)=p_c=8/3$.} 
\label{fig11}
\end{figure}

In Fig.~\ref{fig10} we show the values obtained for the gap $E_1$, for the 
case of open  (Fig.~\ref{fig10}a) and periodic boundary conditions 
(Fig.~\ref{fig10}b). The data are for  lattice 
sizes up to $L=18292$. We see from these figures that the leading 
finite-size scaling behavior of the gap is given by:
\be \label{e4.2}
E_1 \sim \frac{1}{L^z}f(\Delta L^{1/\nu}),
\ee
where $\Delta= p_c-p$, with $p_c=8/3$, $z=0.563$ and $\nu=2.103$. 
 We also see from the figure that the 
scaling function $f(0)\neq 0$ and  decreases exponentially with the 
argument. This imply that the decay time $1/E_1$ of the quasi-stationary 
state increases exponentially as the  lattice size increases. For $p>p_c$ 
we verified that the quasi-stationary state disappears and the system
  is trapped in one of the absorbing states. 
We then expect that 
$p=p_c$ is a critical point. The PARPM with the values $p={p}_1$ and 
$p={p}_c$ seems  to be similar to the bimodal and spinodal limits of 
metastable systems \cite{debenedetti1996metastable}. Considering different values of $u$ for the 
PARPM we should have as the limiting values $p={p}_1=2$ and $p_c(u)$ 
for the region with absorbing sates. 
For $u=0$ we have seen in Sec.~4 that $({p}_1,p_c)= (2,2)$, and in Sec.~6  we are going to see that for $u\to \infty$, $({p}_1,p_c)=(2,4)$.  We can then reach the 
critical point ($2,8/3$) at $u=1$ by considering $u\gtrsim 1$ with the 
parameter $p=8/3$ fixed (see Fig.~\ref{fig0}). In Fig.~\ref{fig11}a this was done for the case of 
periodic boundaries, and we see that, as in Fig.~\ref{fig10}a, the gap behaves 
as in \rf{e4.2}. These results imply that at $u=1$ and $p =p_c=8/3$, the energy 
gap has the $L$-dependence $E_1\sim 1/L^z$, with the dynamical critical 
exponent $z\sim 0.51-0.57$. To improve this estimate we consider a sequence 
of periodic lattices with $712 \leq L\leq 370944$. The results are 
shown in Fig.~\ref{fig11}b and give us $z= 0.561\pm 0.008$.

In summary for $0\leq p <{p}_1$ the model is conformally invariant with 
dynamical critical exponent $z=1$. For ${p}_1\leq p<p_c$ the model 
is in a phase containing multiple absorbing states, but these absorbing states 
are only reached after a time interval that increases exponentially with the 
lattice size. In this phase the system stays in a quasi-stationary state with 
dynamical critical exponent $z=1$ and similar properties as in the conformally 
invariant phase $p<{p}_1$. The sound velocity for $0\leq p \leq p_c$ 
is proportional to $(p-8/3)$. At $p=8/3$ it vanishes and we have a new 
critical behavior with $z=0.561 \pm 0.008$. A similar vanishing of the sound 
velocity happens in the XXZ quantum chain when we approach the ferromagnetic 
point ($z=1$) coming from the anti ferromagnetic conformally invariant 
phase ($z=1$) \cite{ABB}. For values $p>p_c=8/3$ the model is frozen in one of 
the infinitely many absorbing  states.

\section{The model with no desorption ($u\to \infty$)}

We consider in this section the PARPM in the case we have only adsorptions, 
i. e., $u_d=0$ and $u=u_a/u_d \to \infty$. In the stochastic evolution only 
the tilted tiles that hit sites with valleys ($h_{i-1}>h_i<h_{i+1}$) produce 
changes in the growing surface ($h_i \to h_{i} +2$). This model can be 
interpreted as a kind of a single step model \cite{barabasi1995fractal,Sasamoto2010} without 
desorption. 

\begin{figure}
\centering
\includegraphics[angle=0,width=0.4\textwidth] {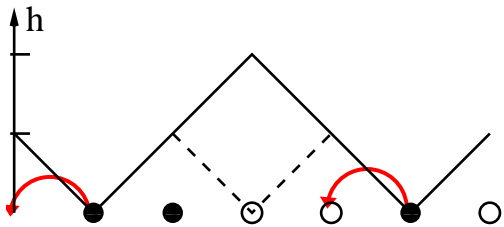}
\caption{
The correspondence among the configurations in the height  and 
the particle-hole representations (circles)  in the periodic PARPM. 
At $u \to \infty$ 
(no desorptions) the 
only allowed motions for the particles are shown in the figure.}
\label{fig12}
\end{figure}

It is interesting to map the  configurations of the PARPM in terms of 
excluded volume particles and holes in the lattice, as in \cite{alcaraz2013nonlocal} 
(particle-hole representation). The 
particles (holes) are defined on the sites $i$ where 
$h_{i+1}>h_i$ ($h_{i+1}<h_i$). In this map the number of particles and holes 
are equal (see Fig.~\ref{fig12}, as an example). The adsorptions in the sites 
with valleys correspond to the motion of a particle to the leftmost 
lattice position,  
provided it is empty (hole). For $u\to \infty$ this is the only allowed 
motion as happens in the totally asymmetric exclusion process (TASEP)
\cite{derrida1993exact,derrida1997exact}. The PARPM with the choice $p=1$ recovers the standard RPM. In 
this case all the sites during the stochastic evolution are chosen with 
equal probability and the particle-hole mapping give us the standard TASEP 
model \cite{derrida1993exact,derrida1997exact}. In the case of open boundary conditions the model is 
noncritical. The stationary state, that corresponds to the pyramid 
configuration in the height  representation is the configuration where the 
$L/2$ particles are in the left half of the $L$-sites lattice. However for 
periodic chains the model is critical and belongs to the 
Kardar-Parisi-Zhang (KPZ) universality class, whose dynamical critical 
exponent $z=3/2$.

For general values of the parameter $p$ the PARPM with no desorption ($u_d=0$) 
and  adsorption rate $u_a$ will be related to a generalized TASEP whose 
transition rates for the particles hoppings depend non locally on the 
particles positions. As a consequence of our definition \rf{e2.6} 
the transition rates depend on the number of peaks (or valleys) in the height 
representation of the model. In the particle-hole representation this is the 
total number  $N_{ph}^c$ of  hole-particle pairs  
(hole in the left and a particle 
in the right), or equivalently is the number of 
particles's clusters in the configuration. From \rf{e2.6p} the transition rates 
for the particles in the generalized TASEP is $u_aq_c$ where 
\be \label{e5.1}
q_c = \mbox{Max}\left\{ 0, \frac{1 -\frac{N_{ph}^c p}{L}}{1-\frac{N_{ph}^c}{L}}\right\},
\ee
where for $p=1$ we have the standard TASEP with transition rate $u_a$.

Since  $N_{ph}^c=L/2$ is the maximum value, for $0\leq p<2$ the parameter $p$ 
acts as a  nonlocal perturbation in the TASEP ($p=1$). When $p>1$ ($p<1$) the 
configurations with  larger (smaller) number of pairs $N_{ph}^c$ are 
preferred. For $p\geq 2$ this perturbation introduces a new effect. At $p=2$ 
the two configurations with $N_{ph}^c=L/2$ (all the particles occupying the 
even or odd sites) become absorbing states. As we increase $p\geq 2$, the 
configurations $c$  with density of pairs $N_{ph}^c/L\geq 1/p$ also become 
absorbing states. For $p>2$ and $L\to \infty$ we have then an infinite number 
of absorbing states in a model with particle number conservation. This is 
similar as happen in  other models like the Manna model, the conserved 
threshold transfer process (CTTP) and in the conserved lattice gas (CLG)
\cite{odor2008universality,henkel2008non}. However, as we shall see, the phase transition to the 
frozen state in our model shows a dynamical critical exponent $z\approx 1$, 
different from the values $z=1.39$ for the Manna model and CTTP and 
$z=2$ for the CLG \cite{odor2008universality,henkel2008non}.

As happened in the case $u=1$ our results indicate that for 
$0\leq p <{p}_1 =2$, where the model has no absorbing states the dynamical 
critical exponent has the same value as in the $p=1$ standard TASEP 
where $z=3/2$. In the region $2< p<p_c$ the model, although having in the 
thermodynamical limit an infinite number of absorbing states, stays in a 
quasi-stationary state with the same critical properties as in the KPZ region  
$0 \leq p<2$, i.e.,  $z=3/2$. The quasi-stationary state for $2\leq p<4$ decays 
in  an absorbing 
state after a transient time that diverges exponentially with the lattice 
size of the system. The net effect of the parameter $p$, for $0\leq p<4$ is
 a change of the sound velocity of the model, that is proportional to 
$4-p$. 
\begin{figure}
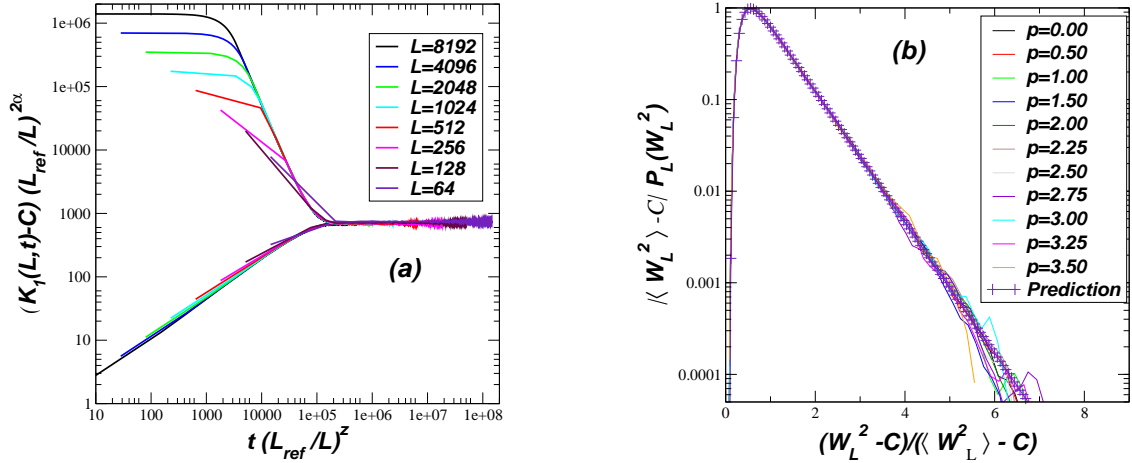

\begin{subfigure}{.5\textwidth}
\centering
\includegraphics[angle=0,width=0.8\textwidth] {fig13a-diego.eps}
\label{fig13a}
\end{subfigure}
\begin{subfigure}{.5\textwidth}
\centering
\includegraphics[angle=0,width=0.8\textwidth] {fig13b-diego.eps}
\label{fig13b}
\end{subfigure}
\caption{
(a)  The time evolution of the first cumulant 
$K_1(L,t)$ for the PARPM at $u\to \infty$ and the 
parameter $p=0.5$, for several 
lattice sizes. The top curves (bottom curves) are the ones where the initial configuration 
is the pyramid (substrate). The constant $L_{ref}=8192$ was introduced 
to better represent the data and $C=0.594$ is obtained as a result of the 
data collapse of the curves, as the time increases, and by fixing the 
KPZ exponents $z=3/2$ and $\alpha=1/2$.
  (b)  Probability distribution of 
the second moment $W_L^2$ of the height distribution at the stationary ($p<{p}_1=2$)  or 
quasi-stationary state ($p>2$) of the PARPM with no desorption 
($u\to \infty$). It is also 
included (crosses) the theoretical KPZ prediction for this distribution \cite{antal2002roughness}.}
\label{fig13}
\end{figure}

In order to identify the universality class of the PARPM for general values 
of $p$  and $u_d=0$ we calculate the time evolution of the surface 
roughness 
\be \label{e5.2}
W_L(t) = \sqrt{\frac{1}{L} \sum_{i=1}^L (h(i,t) - \bar{h}(t))^2},
\ee
where $\bar{h}(t)= \frac{1}{L}\sum_i h(i,t)$ is the average height of the 
configuration at time $t$. In Fig.~\ref{fig13}a we show the time evolution of the first 
cumulant of $W_L^2$, i. e., $K_1(L,t) =<W_L^2(t)>$, for the model with $p=0.5$. 
We consider as the initial conditions the configurations with a single peak 
(top curves) and $\frac{L}{2}$ peaks (bottom  curves). The collapse of the 
curves in the infinite-size  limit 
\be \label{e5.3}
<W_L(t)> \sim L^{\alpha}f(t/L^z),
\ee
give us $z=3/2$ and $\alpha=1/2$, indicates  that the model belongs to  the 
KPZ universality class of critical 
behavior. We also consider, for the lattice size $L=4096$, the probability 
distribution of $W_L^2$ in the stationary states for $p<{p}_1$, and also for 
${p}_1 \leq p <p_c$. This is shown in Fig.~\ref{fig13}b. 
We also include in this last figure  
 the theoretical prediction  for the KPZ model 
\cite{antal2002roughness} (crossed points). We clearly see a nice agreement, indicating that 
indeed for $0\leq p < p_c$ the model belongs to the KPZ universality class, 
where $z=3/2$ and $\alpha=1/2$. For all cases the probability distribution 
behaves as 
\be \label{e5.4}
P_L(W_L^2) = \frac{1}{<W_L^2> -C } f\left(\frac{W_L^2-C}{<W_L^2>-C}\right), 
\ee
where $C=0.594$ is a finite-size correction and $<\cdot >$ is the standard 
average 
over the configurations \cite{oliveira2007finite}.
\begin{figure}
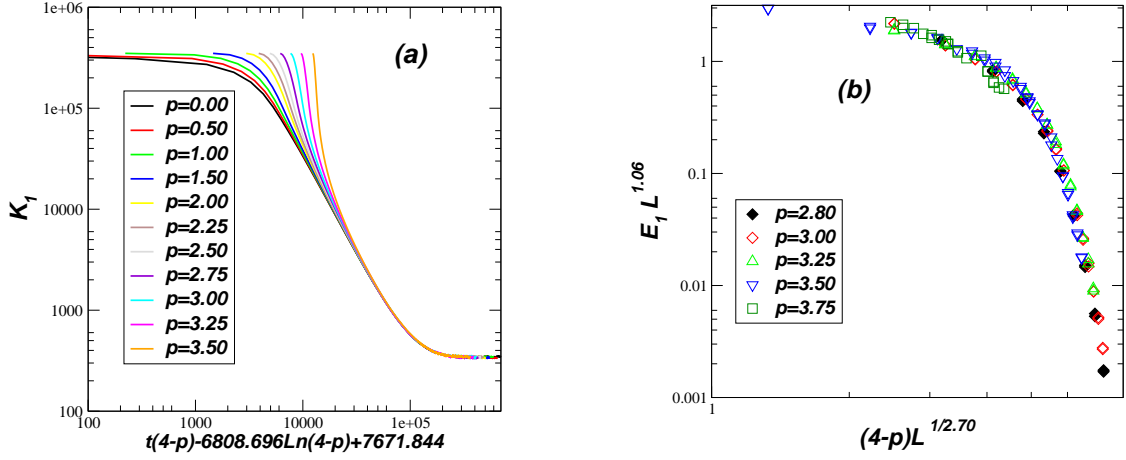

\begin{subfigure}{.5\textwidth}
\centering
\includegraphics[angle=0,width=0.8\textwidth] {fig14a-diego.eps}
\end{subfigure}
\begin{subfigure}{.5\textwidth}
\centering
\includegraphics[angle=0,width=0.8\textwidth] {fig14b-diego.eps}
\label{fig14b}
\end{subfigure}
\caption{
(a)  The time evolution of the first cumulant 
$K_1(L,t)=<W_L^2>$ for the PARPM at $u\to \infty$ with
several values of the parameter $p$ (shown in the figure) and  for the 
lattice size $L=4096$. As time grows the curves collapse if  the time 
is scaled by the factor $4-p$. 
  (b)  Finite-size scaling for the gap of the PARPM at $u \to \infty$, for 
several values of $p<4$. The gaps were estimated by fitting ($1-\tau/\tau_{\infty}$) with \rf{e4.2} and $\Delta=4-p$. }
\label{fig14}
\end{figure}

We want now to consider the limiting case $p=p_c=4$. In Fig.~\ref{fig14}a 
we show the time evolution, for several values of $p<p_c$, of the average 
$K_1(L,t)$ for the lattice size $L=4096$. We see from the figure that the 
time scale changes as a function of $p-4$. We also notice  
the existence of a transient regime, that increases as we get closer to 
$p_c$. This is an indication that at $p_c$ the system may have a distinct 
behavior. In order to verify this possibility we need to evaluate the 
lowest gap $E_1$ for $p<p_c$. As we did in Sec. 4 for the case $u=1$, this 
gap could be calculated from the large-time dependence of the observable 
$\tau_L(t)$ (see \rf{e4.1}). In Fig.~\ref{fig14}b we show the 
finite-size scaling obtained for the gap at $u \to \infty$. 

\begin{figure}
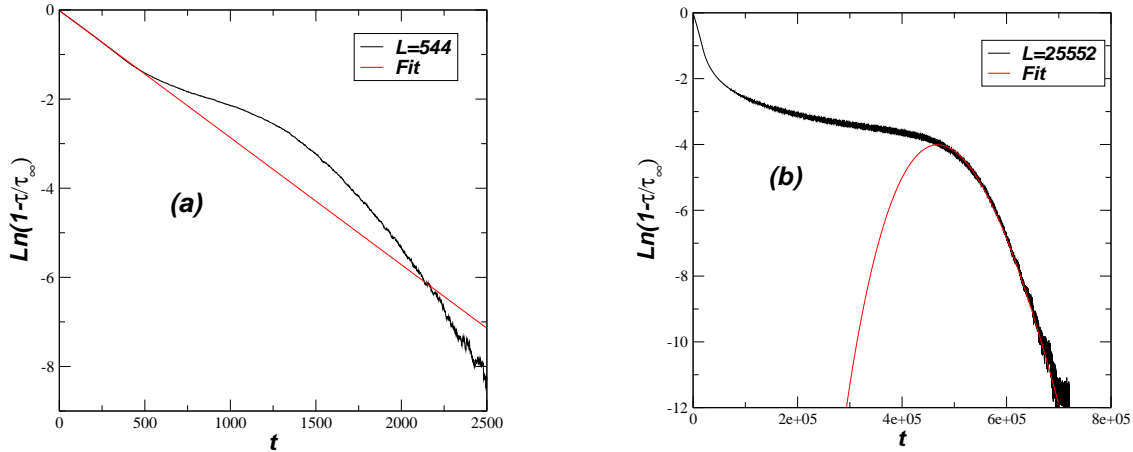

\begin{subfigure}{.5\textwidth}
\centering
\includegraphics[angle=0,width=0.8\textwidth] {fig15a-diego.eps}
\label{fig15a}
\end{subfigure}
\begin{subfigure}{.5\textwidth}
\centering
\includegraphics[angle=0,width=0.8\textwidth] {fig15b-diego.eps}
\label{fig15b}
\end{subfigure}
\caption{(a)  Semi-log plot of the ratio 
$\tau_L(t)/\tau_L(\infty)$ (black curve),as a function of time, 
for the periodic PARPM 
with $L=544$ sites,  no desorptions ($u\to \infty$) and 
at the critical point $p_c=4$. The initial state 
of the system is  the pyramid configuration.  The red line is the linear 
fit obtained for $t<500$, as expected from \rf{e4.1}. (b) Same as in  (a) with the larger  
lattice size $L=25552$. The curve in red is obtained by fitting, according 
to \rf{e5.5}, the data for the larger  times 
 $t> 5\times 10^5$.}
\label{fig15}
\end{figure}

\begin{figure}
\centering
\includegraphics[angle=0,width=0.4\textwidth] {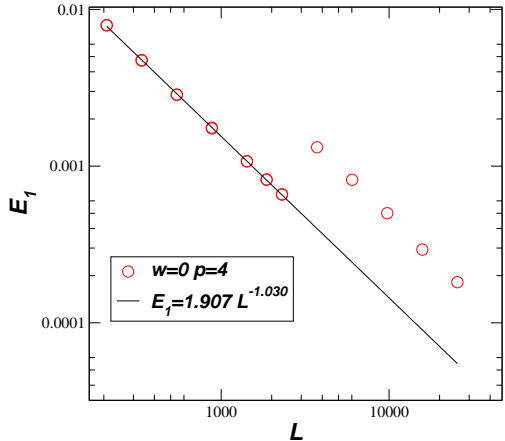}
\caption{
The gap estimates obtained from the time dependence of 
$\ln[1-\tau_L(t)/\tau_L(\infty)]$ of the PARPM with periodic boundaries, 
$u\to \infty$ and $p=p_c=4$.
For $t < 1500$
  the estimates were obtained from the 
fitting with \rf{e4.1} and for large times the fitting was done by using  
\rf{e5.5}.} 
\label{fig16}
\end{figure}

In Fig.~\ref{fig15}a we show in a semi-log plot 
the quantity $1-\tau_L(t)/\tau_L(\infty)$, starting  with the system in the 
pyramid configuration. We clearly see from this figure that the exponential 
time dependence \rf{e4.1} is only obtained for short times. As we obtained 
in the $u=0$ case (see Sec.~3), if the eigenvalue $E_1$ has an associated 
Jordan-cell structure we may have instead of a simple exponential decay as 
in \rf{e4.1}, the more general time dependence 
\be \label{e5.5} 
1-\frac{\tau_L(t)}{\tau_L(\infty)} \sim t^n \exp (-E_1t),
\ee
where $n$ is an $L$-dependent integer and $E_1$ is the gap. In 
Fig.~\ref{fig15}b we fit, using \rf{e5.5}, the results obtained for the lattice 
$L=25552$. We obtained a good fitting for $ t > 5\times 10^5  t$
 giving us the value $E_1=1.81\times10^{-4}$ and 
$n=85$. Repeating the same fitting for the lattice sizes 
$L=3728$,  $6032$, $9760$ and $15792$  we obtain  
$E_1\sim 1/(0.22L - 87.81)$ and $n\sim 0.213L^{0.59}$. This result  indicates 
that indeed at $p=p_c$ we have a new critical behavior with the dynamical 
critical exponent $z=1$. From Fig.~\ref{fig15}a we see that the simple 
exponential as in \rf{e4.1} give us the short-time behavior. Extracting 
the gap from the initial times, assuming \rf{e4.1} for the small lattices 
$208 \leq L\leq 2304$ we obtain $E_1 \sim 1/(0.66L-11.07)$. These results 
together with the ones obtained by using the fitting \rf{e5.5} for large times 
are shown in Fig.~\ref{fig16}.  There is however an apparent contradiction 
in these results since the gap calculated from  the large time decay is 
greater than the one governing the short-time behavior. This is an indication 
that indeed at large times the lowest gap $E_1$ appearing in \rf{e5.5} should 
be replaced by a higher excited state $E_i>E_1$. This means that the 
Jordan-cell structure dominating the $n$-polynomial behavior at large times 
is not associated to the first excited state, but to a higher excited state. 
The large dimension of the Jordan cells associated to $E_i$ are enough to 
kill the exponential decays associated to the smaller gaps. 

In summary for $u=\infty$ the PARPM is in the KPZ universality class 
($z=3/2,\alpha=1/2$) for all the values of $0\leq p<p_c=4$. For $2< p <p_c$, 
although there exist an infinite number of absorbing states, the system stays 
during a time interval that grow exponentially with the lattice size, in 
a quasi-stationary state sharing the  critical properties  of the 
 KPZ universality class. For $0\leq p <p_c$ 
the large-time behavior has a simple exponential decay with a time decay 
$1/E_1$. However at $p=p_c=4$ our results indicate that a Jordan-cell 
structure associate to a higher gap $E_i>E_1$, dominates the large-time 
behavior. In this case the system stays in a quasi-stationary state with a 
dynamical critical exponent $z\approx 1$. For $p>p_c$ the model is in a frozen state, 
i.e., after a short time  decays to one of the infinitely many absorbing states. 

\section{Conclusions}

  The peak adjusted raise and peel model (PARPM) was studied in this paper. 
The model can describe the interface fluctuations of a growth model (height 
representation) or the particle density fluctuations in a nonlocal 
particle-hole exclusion process (particle-hole representation). The model 
has two parameters: $u$ and $p$. The parameter $u\geq 0$ controls the ratio
 between the rates of adsorption and desorption in the height representation, 
and in the particle-hole representation it gives the relative weight among the 
local and nonlocal hoppings of the particles. The parameter $p\geq 0$ 
distinguish, during the time evolution, the configurations according to the 
number of peaks, in the height representation, or the number of 
particle-hole pairs in the particle-hole representation.  
In the bulk limit ($L\to \infty$), for any value of $u$ and  $p>2$,  
the model has an infinite number 
of absorbing states. An schematic phase diagram of the model is shown in 
 \ref{fig0}. We study the model with open and periodic boundary 
conditions by choosing some special values of $u$ and general values of $p$.

{\it a}) At $u=0$ (Sec.~3) where we have no desorption in the height 
representation, the model is gapped for $p<2$, having an absorbing state as 
the stationary state. In the bulk limit the associated Hamiltonian has an 
infinite dimensional  Jordan cell structure associated to the first excited 
state. Starting from some simple configurations we were able to derive the 
analytical time dependence of some observables for arbitrary time. 
Interestingly, if  we consider in the $L \to \infty$, a sequence of initial 
configurations where the density of tiles in the first row  
is fixed and nonzero, the system 
stays an infinite time before decaying  into the final absorbing state 
and,  does not 
fell the system's gap. At $p=2$ however, the model is critical with a 
dynamical critical exponent $z=1$. For $p>2$ the stationary state is one of 
the infinitely many absorbing states. 

{\it b}) For $u=1$ we extend the results obtained in \cite{alcaraz2010conformal,alcaraz2011conformal} for 
$p\leq 2$, where the model is known to be conformally invariant 
(central charge $c=0$ and $z=1$). Our results show that for $2\leq p <8/3$, 
where the model has an infinite number of absorbing states, it stays 
in a critical quasi-stationary state during a time interval that grows 
exponentially with the lattice size. The quasi-stationary state has the same 
critical properties as the conformally invariant phase $p<2$. At $p=p_c=8/3$ 
our results give us a new critical behavior with dynamical critical exponent 
$z\sim 0.561$. For $p>p_c=8/3$ the system, after a short time, stays 
inactive in one of the infinitely many absorbing states. The phase transition at 
$p_c=8/3$ separates an active phase from an inactive phase of a system with 
an infinite number of absorbing states. We think that is the first time we see 
such a phase transition, i.e., a transition to a frozen inactive state 
coming from a critical phase, that although having infinitely many 
 absorbing states, is ruled by an active quasi-stationary state. 

{\it c}) For $u\to \infty$ the model has no desorption and it loses part 
of its non locality, since the adsorptions are local. In the particle-hole 
representation  the model is equivalent to a TASEP where we have $L/2$ 
particles that can hop to the next-neighbor site at the left, provided it is 
empty (hole). For a given value of the parameter $p$ the rate of the 
hopping process depends on the number of particle-hole pairs in the 
configuration. In this limit the model is gapped or not depending if the 
boundary condition is open or periodic, respectively. At $p=1$ the motion of 
the particles are independent of the numbers of particle-hole pairs and the
model recovers the standard critical TASEP with the dynamical critical 
exponent $z=3/2$ of the KPZ universality class. Our results  indicate that 
for all values $0\leq p < p_c=4$, the model has the same critical behavior 
as the KPZ universality class. For $2 \leq p<p_c=4$, similarly as happened in 
the case $u=1$, the model although having an infinite number of absorbing 
states, survives for a time interval that diverges exponentially with the 
lattice size, in a critical quasi-stationary state whose critical properties  
are the same as the ones on the region without absorbing states 
($0 \leq p <2$). At $p=p_c=4$,  as in the cases $u=0$ and $u=1$, the 
model shows a new critical behavior with dynamical critical exponent $z=1$. 
We also noticed that this new critical behavior is  a consequence of large 
 dimensional Jordan cells associated not to the first excited state, but to a 
higher excited one. 

To conclude it is interesting to see the PARPM as a generalized asymmetric 
exclusion process ASEP \cite{derrida1997exact}, with nonlocal jumps of excluded volume 
particles (controlled by $u$) whose rates are biased (controlled by $p$) by 
the number of particle-hole pairs in the configurations. The parameter $u$ 
is relevant since it changes the critical behavior. The parameter $p$, on the 
other hand, is irrelevant up to $p=p_c(u)$ 
($p_c(0) =2, p_c(1)=8/3, p_c(\infty)=4$), although for $2\leq p \leq p_c(u)$ 
there exist an infinite number of absorbing states. At $p=p_c(u)$ the model
has a new critical behavior with a $u$-dependent critical exponent 
$z=z(u)$.

\section{Acknowledgments}
We would like to thank Jos\'e Abel Hoyos and Vladimir Rittenberg for useful discussions and a careful reading of the manuscript.  This work was support in part by the Brazilian funding agencies: 
  FAPESP, CNPq and CAPES.

\end{document}